\pdfoutput=1

\documentclass[%
 aip,
 amsmath,amssymb,
 reprint,%
]{revtex4-1}

\usepackage{graphicx}
\usepackage{dcolumn}
\usepackage{bm}

\usepackage[utf8]{inputenc}
\usepackage[T1]{fontenc}
\usepackage{mathptmx}
\usepackage{amsmath}
\usepackage{textcomp, gensymb}
\usepackage{etoolbox}
\usepackage{siunitx}
\usepackage{upgreek}
\usepackage{xcolor}
\usepackage{soul}

\makeatletter
\def\@email#1#2{%
 \endgroup
 \patchcmd{\titleblock@produce}
  {\frontmatter@RRAPformat}
  {\frontmatter@RRAPformat{\produce@RRAP{*#1\href{mailto:#2}{#2}}}\frontmatter@RRAPformat}
  {}{}
}%
\makeatother

\begin{document}

\preprint{AIP/123-QED}

\title{Tailoring Nanowire Lasing Modes via Coupling to Metal Gratings }
\author{F. Vitale}
\affiliation{ 
Institute of Solid State Phyisics, Friedrich Schiller University Jena, Max-Wien-Platz 1, 07743 Jena, Germany
}%
 \email{francesco.vitale@uni-jena.de}
\author{D. Repp}%
\affiliation{ 
Institute of Applied Physics, Abbe Center of Photonics,Friedrich Schiller University Jena, Albert-Einstein-Straße 15, 07745 Jena, Germany
}
\author{T. Siefke}%
\affiliation{ 
Institute of Applied Physics, Abbe Center of Photonics,Friedrich Schiller University Jena, Albert-Einstein-Straße 15, 07745 Jena, Germany
}
\author{U. Zeitner}%
\affiliation{ 
Institute of Applied Physics, Abbe Center of Photonics,Friedrich Schiller University Jena, Albert-Einstein-Straße 15, 07745 Jena, Germany
}
\affiliation{Fraunhofer Institute for Applied Optics and Precision Engineering, Albert-Einstein-Straße 7, 07745 Jena, Germany
}
\author{U. Peschel}%
\affiliation{%
Institute of Condensed Matter Theory and Solid State Optics, Friedrich Schiller University Jena, Max-Wien-Platz 1, Jena 07743, Germany
}%
\author{T. Pertsch}%
\affiliation{ 
Institute of Applied Physics, Abbe Center of Photonics,Friedrich Schiller University Jena, Albert-Einstein-Straße 15, 07745 Jena, Germany
}
\author{C. Ronning}
\affiliation{ 
Institute of Solid State Phyisics, Friedrich Schiller University Jena, Max-Wien-Platz 1, 07743 Jena, Germany
}%

\date{\today}

\begin{abstract}
Tailoring the emission of plasmonic nanowire-based lasers represents one of the major challenges in the field of nanoplasmonics, given the envisaged integration of such devices into on-chip all-optical circuits. In this study, we proposed a mode selection scheme based on distributed feedback, achieved via the external coupling of single zinc oxide nanowires to an aluminum grating, which enabled a quasi-single mode lasing action. The nano-manipulation of a single nanowire allowed for a reliable comparison of the lasing emission characteristics in both planar (i.e. nanowire on the metallic substrate) and on-grating configurations. We found that, by varying the orientation of the nanowire on the grating, only when the nano-cavity was perpendicular to the ridge direction, an additional peak emerged in the emission spectrum on the low-energy side of the gain envelope. As a consequence of the fulfillment of the Bragg condition, such a peak was attributed to a hybrid mode dominating the mode competition. Simulation results showed that the hybrid mode could be efficiently waveguided along the nanowire cavity and supported by localized plasmon polaritons building up at the raised features ("fences") on top of the metal grating ridges. Moreover, the hybrid mode was found to experience an extra reflectance of nearly 50\% across the grating periods, in addition to that provided by the nanowire end facets.
\end{abstract}

\maketitle

Since the theoretical formulation of the SPASER \cite{bergman2003surface} and following the first experimental realization \cite{oulton2008hybrid, oulton2009plasmon}, nanowire-based plasmonic lasers - mostly designed as a semiconductor-insulator-metal (SIM) scheme - have gained an ever-increasing scientific appeal \cite{lu2012plasmonic, chou2016high, chu2022dimension}. In particular, due to the buildup of surface plasmon polaritons (SPPs) at the interface between the nanowire and the metal film, plasmonic nanolasers can overcome the diffraction limit via the sub-wavelength confinement of the modal field \cite{ma2011room}. As a consequence, such devices are characterized by smaller modal volumes \cite{bergman2003surface, oulton2012surface}, a larger Purcell-enhanced stimulated emission \cite{zhang2014room, chou2015ultrastrong} and, thus, a faster lasing dynamics \cite{stockman2010spaser, sidiropoulos2014ultrafast}, when compared to photonic nanolasers \cite{lu2017plasmon, wille2016carrier}. However, within the framework of next-generation all-optical circuitry \cite{de2012quantum, davis2017plasmonic, fukuda2020feasibility}, the tailoring of the emission characteristics and the lowering of the lasing threshold still represent the major challenges in the field for the integration of plasmonic nanolasers \cite{xu2019surface, liang2020plasmonic}. Along with improving the metal quality and smoothening the surface of the metallic substrate \cite{yu2017influence, chou2016single, liao2019low}, researchers have started focusing on the modification of the plasmonic cavities to boost their lasing performances, either by structuring the dielectric/metal substrate \cite{bermudez2017plasmonic, chou2018ultracompact, cheng2018epitaxial} or by hybridizing the dielectric spacer between the nanowire and the metal film \cite{li2019plasmonic, li2020current}. \newline
In this work, we made use of a mechanical nano-manipulation approach, to study the effects of a metal grating on the lasing properties of semiconductor nanowire-based plasmonic lasers. Namely, the same nanowire could be manipulated and “switched” among different configurations: from placed on the dielectric/metal substrate (planar plasmonic configuration) to overlaid on a metal grating (nanowire-metal grating configuration) at different orientations with respect to the ridge (or trench) direction. To ensure reliable comparability among the different configurations, the single-nanowire experiments compensated for the statistical fluctuations arising from the wire-to-wire variations in crystal quality, cavity size (length and diameter), and morphology (if tapered, strained, roughened or cleaved at the end facets): all parameters that significantly affect the cavity Q factor and the overall emission characteristics \cite{zimmler2010optically, couteau2015nanowire, zapf2019tailoring}. To this end, we used Al gratings characterized by geometrical parameters tailored to the lasing emission wavelength, in order to externally induce additional distributed feedback to the gain spectrum. \newline 
Zinc oxide (ZnO) nanowires were produced by an Au-catalyzed vapor-liquid-solid (VLS) growth inside a horizontal three-zone tube furnace: further details can be found in reference \cite{borchers2006catalyst}. Afterward, the nanowires were dry-imprinted from the growth sample onto the dielectric/metal substrates. Such substrates consisted of a native Al$_2$O$_3$ layer (thickness $t\approx $ 3 nm), serving as a spacer for enhancing the modal field confinement and for mitigating the metal losses, on top of a polycrystalline Al layer ($t\approx$ 70 nm), which featured the co-deposition of 10 at\% Si for decreasing the surface roughness. Subsequently, these substrates were subdivided into etching spots that were patterned via electron beam lithography (EBL) for the fabrication of the metal gratings \cite{kley2012enhanced}: for further details see the Supporting Material, section S1. \newline
The optical characterization of single nanowires in a given configuration was accomplished by focusing the emission from a frequency-tripled Nd:YAG laser ($\lambda_{\textup{exc}}$ = 355 nm, $f_{\textup{rep}}$ = 100 Hz, $t_{\textup{pulse}}$ = 7 ns) through a 50x NUV objective (NA = 0.43), down to a spot size of about 20 $\upmu$m, to encompass the whole nanowire. The spectral acquisition was carried out by using a spectrometer (Princeton Instruments SP-2500i) equipped with a 1200 lines/mm grating (blazed at 300 nm) and connected to a liquid-nitrogen-cooled, front-illuminated CCD camera. The photoluminescence measurements were performed at a temperature of 150 K inside a liquid He flow cryostat (Janis ST-500). 

\begin{figure} [ht!]
    \includegraphics[width=0.75\columnwidth]{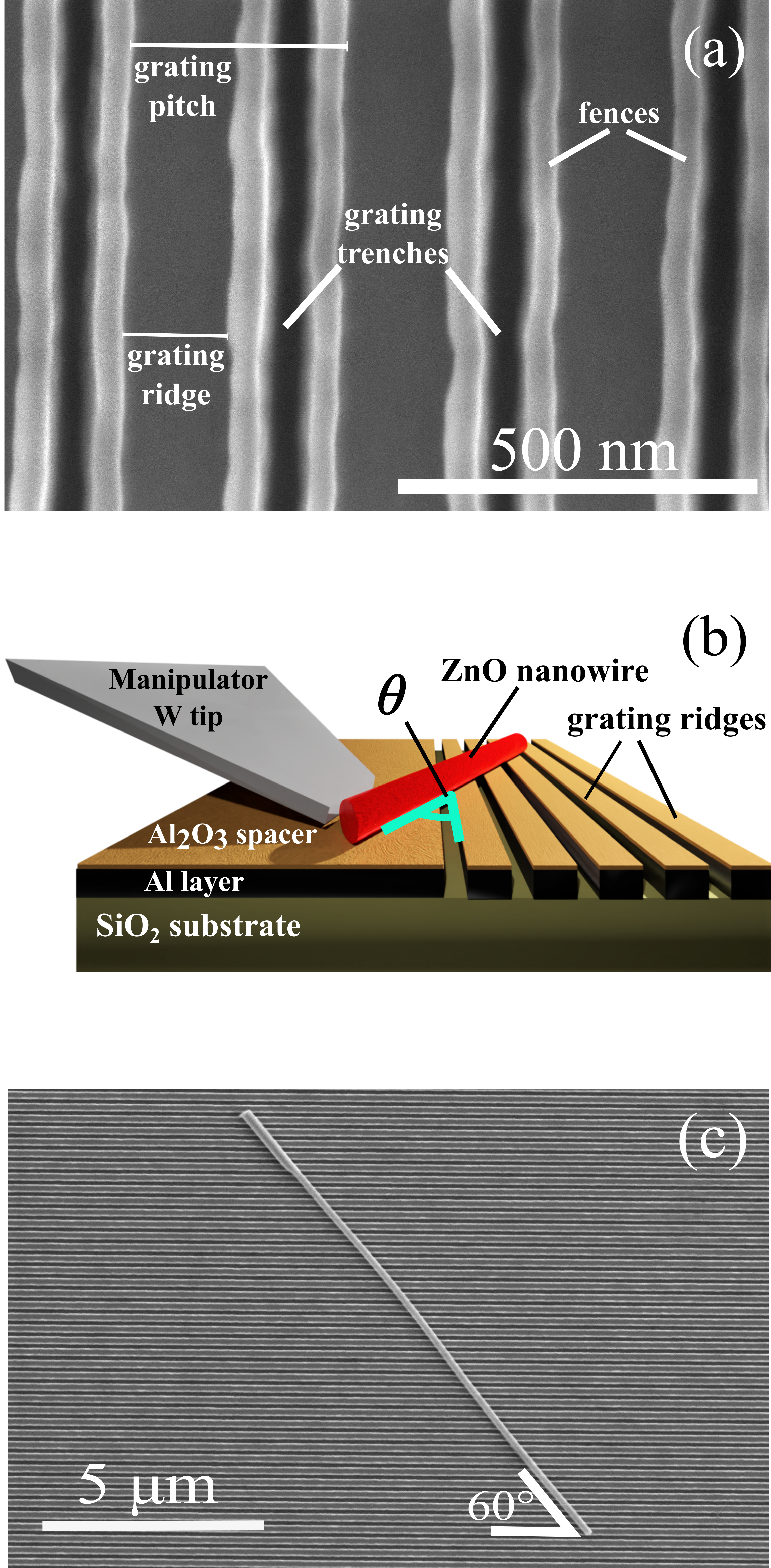}
        \caption{(a) SEM image of a “fenced” Al grating, featuring a pitch $p$ = 300 nm and fill factor (i.e. the trench-width/pitch ratio) $FF$ = 0.3, used for the coupling with a single nanowire. (b) Sketch of the nano-manipulation performed in air: after the characterization on the bare dielectric/metallic substrate, the nanowire was moved onto the metal grating, and oriented at a given angle $\theta$ with respect to the ridge (trench) direction. (c) Exemplary SEM image of a nanowire used in one of the single-nanowire experiments, which was eventually oriented at $\theta \approx 60\degree$ via the nano-manipulation.
}
        \label{Fig.1}
\end{figure}

We started our characterization by screening the randomly distributed nanowires with respect to their morphology (heavily tapered, bundled, and inhomogeneous nanowires were discarded) and proximity to the metal gratings: suitable nanowires were pre-characterized via low-power PL measurements. The nanowire manipulation had to be performed in air under the PL microscope rather than in a scanning electron microscope (SEM), in order to avoid the modification of the optical properties induced by the electron beam \cite{zapf2021performance, lahnemann2016quenching}. Therefore, SEM investigations were performed after all the optical characterizations of each cavity configuration. \newline
Figure \ref{Fig.1}(a) exemplarily shows a top-view SEM image of an Al grating with pitch $p$ = 300 nm, along with the additional “fence-like” structures on top of the metallic ridges. Such structures originated from the pattern transfer processes using ion beam etching (for more details see Supporting Material, section S1). The nanowires were overlaid onto such structures via nano-manipulation, as schematically shown in Figure \ref{Fig.1}(b), which was accomplished by exploiting a single piezo-controlled nano-manipulator (Kleindiek GmbH) equipped with a tungsten tip (further details can be found in the Supporting Material, section S2). Single nanowires were accurately placed on the metal grating and oriented at a certain angle $\theta$ with respect to the ridge direction with an uncertainty of $\approx \pm 5\degree$: the result can be seen in the SEM image of Figure \ref{Fig.1}(c), which exemplarily shows a nanowire used in the PL experiments.

\begin{figure*} [ht!]
    \includegraphics[width=2\columnwidth]{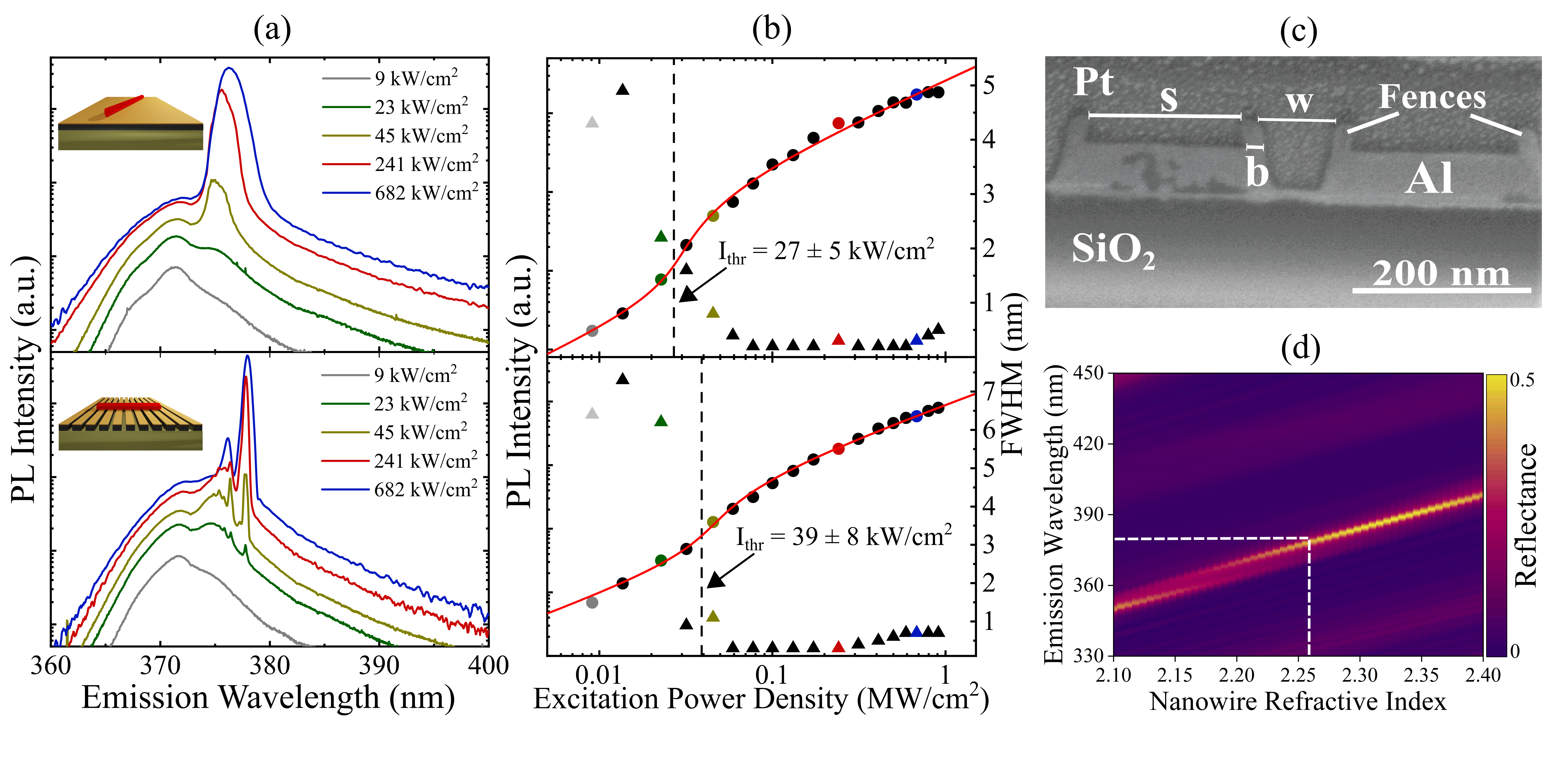}
        \caption{(a) PL spectra of the planar (top panel) and nanowire-grating (bottom panel) configurations acquired for different pump power densities at 150 K; the NW was oriented perpendicular to the ridge direction $\theta \approx 90\degree$. (b) Corresponding light-in-light-out curves (circles) of the emission intensity and spectral width trends (triangles) of the emission for the planar (top panel) and nanowire-grating (bottom panel) configurations, as a function of the excitation power density. The dashed lines indicate the respective lasing thresholds estimated by fitting the experimental data points with an adapted multimode lasing model. (c)  SEM cross-section of an Al grating with a nominal pitch $p$ = 300 nm and fill factor $FF$ = 0.3. The extracted geometrical parameters for the numerical simulations, within the fabrication tolerances and measurement errors, were: fence separation $s$ = 170 nm, fence width $b$ = 20 nm, and trench width $w$ = 90 nm. (d) Simulated reflectance of the “fenced” Al grating as a function of the wavelength and nanowire index. The intersection point of the dashed white lines indicates the reflectance at $\lambda$ = 378 nm for the corresponding material index value ($n_{\textup{ZnO}}$ = 2.26).
}
        \label{Fig.2}
\end{figure*}

Figure \ref{Fig.2}(a) shows the photoluminescence spectra acquired as a function of the excitation power density, with the pump polarization kept perpendicular to the nanowire axis, at a temperature of 150 K, which turned out to be the upper-temperature limit for ensuring good device stability throughout the whole experiment. The spectra were recorded for a ZnO nanowire with length $L \approx$ 12 $\upmu$m and average diameter $\langle d \rangle \approx$ 200 nm, first lying planar on the plasmonic substrate (top panel), and subsequently moved onto the Al grating with an orientation of $\theta \approx 90\degree$ with respect to the ridge direction (bottom panel). The diameter was averaged over the width variations ($\approx \pm 20-30$ nm) along the nanowire, while the widening, usually occurring at one of the end facets as a consequence of tapering, was not considered in the averaged sum. The PL spectrum of the nanowire on the grating clearly shows the emergence of an additional peak on the low-energy side of the gain envelope at $\lambda$ = 378 nm, when compared to the case of the planar configuration. \newline
The geometrical parameters were extracted from the SEM cross-section, shown in Figure \ref{Fig.2}(c), and the structure was reproduced in the commercial FDTD software suite Lumerical \cite{lumerical2019} to calculate the corresponding reflectance spectrum, shown in Figure \ref{Fig.2}(d), as a function of wavelength and refractive index. We found that this structure supports reflectance resonances within the ZnO gain spectrum around 380 nm, where we assumed the refractive index of the nanowire to be $n_{\textup{ZnO}}$ = 2.26, as marked by the white dashed lines in Figure \ref{Fig.2}(d).  
Thus, the emergence of the additional peak can be explained based on the reflectivity characterizing different waveguided modes: indeed, we observed a sharp reflectance peak for a mode polarized perpendicularly to the grating plane. Concerning this mode, we varied the material index until the reflectance peak coincided spectrally with that observed in the experiment, and verified that the Bragg condition:

\begin{equation}
    2 \times \frac{2\pi}{\lambda}n_{eff} = m \frac{2\pi}{p},
\end{equation}

 - where $\lambda$ is the peak wavelength, $n_{\textup{eff}}$ the effective refractive index of the mode, and $p$ the grating pitch -, is fulfilled for a grating order of $m$ = 3. \newline
Figure \ref{Fig.2}(b) shows the corresponding dependence of the photoluminescence intensity on the pump intensity together with that of the full-width-half-maximum (FWHM) of the dominant lasing peak of the nanowire, on the planar substrate (top panel) and on the metal grating (bottom panel). The light-in-light-out curves demonstrate solid device robustness over a large pumping range (about 2 orders of magnitude), with no sign of optical degradation, which could affect the comparison between the two different configurations. These curves were fitted using an adapted multimode lasing model \cite{casperson1975threshold}, which returned as fitting parameters the lasing threshold $I_{\textup{th}}$ and the spontaneous emission factor $\beta$. We found $I_{\textup{th}} = 27 \pm 5$ kW/cm$^2$ and $\beta = 0.06 \pm 0.01$ for the planar configuration and $I_{\textup{th}}= 39 \pm 8$ kW/cm$^2$ and $\beta = 0.10 \pm 0.02 $ for the on-grating configuration, respectively. These fitted values for the lasing threshold also coincide with the characteristic kink observed in the respective “S-shaped” curves, marking the transition from the amplified spontaneous emission to the lasing regime. It is apparent from Figure \ref{Fig.2}(b) that the values for the two configurations are comparable within the experimental errors. From the previously-reported calculated reflectance values, which turned out to be higher than the ones characteristic of the pristine nanowire end-facet cavity \cite{zimmler2010optically}, we expected a lower threshold for the lasing of the nanowire on the grating. However, one should take into account that the peak associated with the mode dominating the lasing action, appears at the low-energy side at $\lambda \approx$ 378 nm, where the gain of ZnO is significantly lower. Furthermore, we attribute the comparable threshold values also to the fabrication imperfections, also visible in Figure \ref{Fig.1}(b), resulting in additional losses at the interfaces between the nanowire and the grating \cite{wright2014distributed}. 

\begin{figure*} [ht!]
    \includegraphics[width=2\columnwidth]{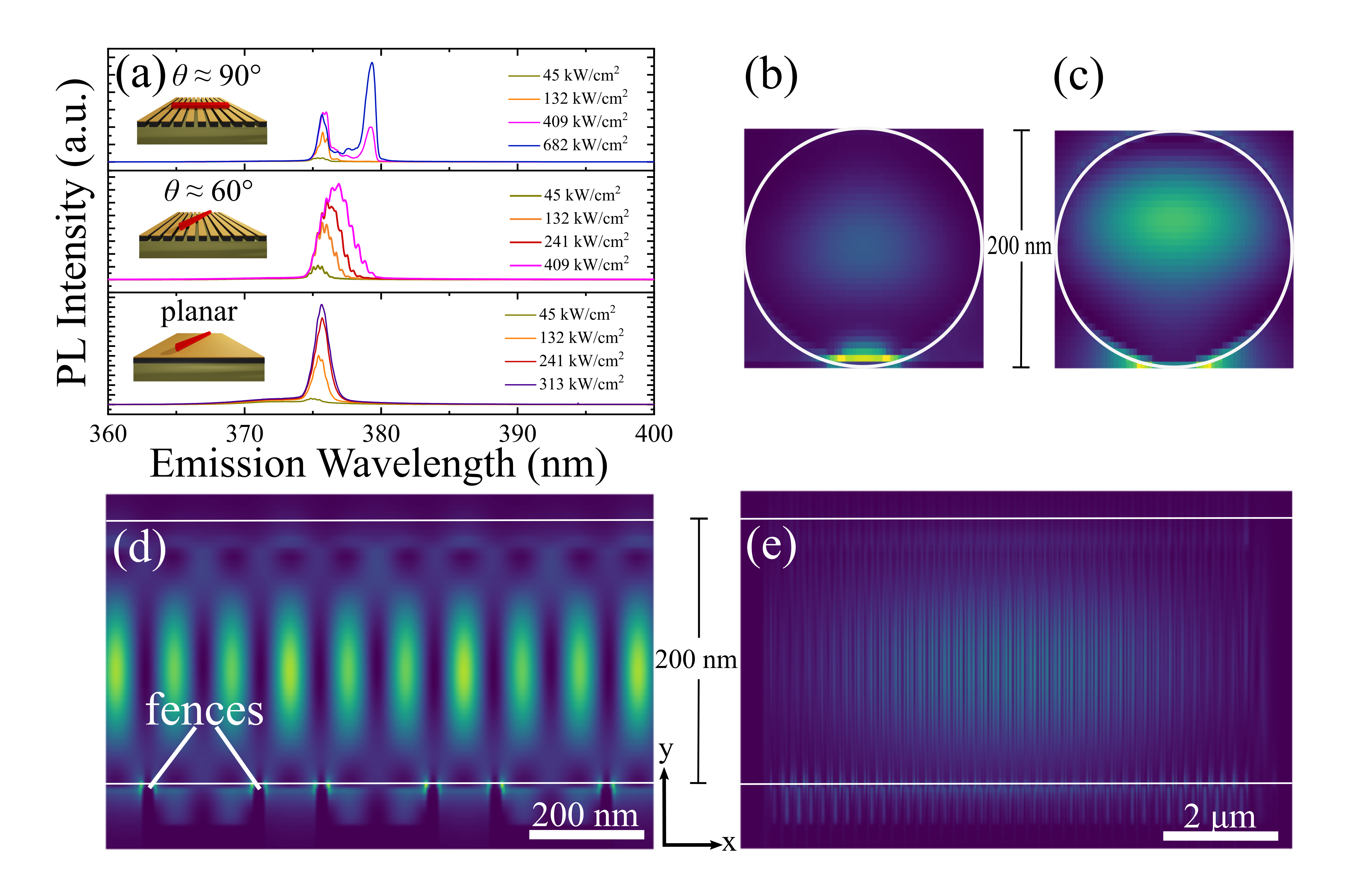}
        \caption{(a) PL spectra of a single ZnO nanowire placed on an Al grating with $p$ = 300 nm and $FF$ = 0.3 and switched between two different orientations ($\theta \approx 90\degree$ and $\theta \approx 60\degree$), then moved onto the metallic substrate, as illustrated by the sketches in the insets. Simulated field distributions of the lasing mode for the experimentally investigated geometry across the nanowire at the (b) fence and (c) free-standing positions. Field intensity distributions of the lasing cavity mode - cut perpendicular to the grating - along the nanowire axis in (d) zoomed in and (e) full view. The  scale bars on the x and y axes are chosen differently for display purposes, while the white solid lines delimit the nanowire diameter. The intensity increase in the mode distribution at the center of the nanowire, when compared to the two end facets, visible in (e), can be likely attributed to the field enhancement given by the metallic fences acting as nano-resonators. As evidenced in (d), 3 maxima appear across one grating period, further corroborating our hypothesis of a third-order distributed feedback mechanism.
}
        \label{Fig.3}
\end{figure*}

To further verify the effects of the extra feedback provided by the metal grating, we conducted other single-nanowire experiments (see also Supporting Material, section S3), featuring an additional re-orientation of the ZnO nanowire on top of the Al grating. Figure \ref{Fig.3}(a) shows the lasing spectra of a nanowire, with cavity dimensions $L \approx 13$ $\upmu$m and $\langle d \rangle \approx$ 250 nm, which was first measured on the Al grating with a ridge orientation of $\theta \approx 90\degree$, then re-oriented at $\theta \approx 60\degree$ and, ultimately, moved off the grating onto the planar substrate. This experiment further corroborates the fact that the appearing emission peak in the low-energy side of the gain spectrum, is caused by the external resonator, which provides additional distributed feedback to the nanowire cavity. Indeed, compared to the other single-nanowire experiment, not only the emission spectrum changes between the planar and grating configurations, but also the further re-orientation at $\theta \approx 60\degree$ leads to the disappearance of the additional peak.
This proves that only for an efficient coupling of the whole nanowire cavity with the fence-like structures, i.e. for an effective pitch matching the coupling length between the nanowire and the metal grating, a “grating-supported” mode starts to dominate the lasing action in the high power regime. \newline

To quantify and obtain more detailed insights into the feedback provided by the coupling of the ZnO nanowires with the metal grating, we performed active lasing simulations in Lumerical (for details see Supporting Material, section S4) by making use of the same above-mentioned parameters. The results show that the excited modes, polarized perpendicular to the substrate plane, attain a field enhancement at the edges of the metallic fences, thanks to the buildup of localized plasmon polaritons \cite{chou2018ultracompact}. This, in turn, leads to a large modal mismatch between the waveguided modes for the free-standing parts of the nanowire, shown in Figure \ref{Fig.3}(b), and those in contact with the metal, shown in Figure \ref{Fig.3}(c). Hence, this results in reflectance values of about 50\%, which add up to the overall end facet reflectivity even for a small number of periods ($N$ = 40), as already shown in Figure \ref{Fig.2}(d). The resulting increase of the quality factor of the cavity mode, thus, causes a considerable line narrowing compared with the nanowire cavity when lying on the planar dielectric/metal substrate.
Indeed, the mode dominating the lasing action is given by the fenced grating rather than by the Fabry-Pérot modes building up between the end facets of the nanowire. Furthermore, one can see that the field intensity distribution along the nanowire is highest in the center of the nanowire, due to the larger and symmetric feedback of the external grating cavity, conversely to the poorer confinement toward the end facets, where one would expect most of the radiation to be coupled out in the absence of the additional grating-supported resonator. This also suggests that an optimal coupling between the nanowire and the grating could be attained only for an orientation of the nanowire perpendicular to the metal grating. Ultimately, due to the action of the grating, the rather broad-band plasmonic lasing emission was converted into narrow-line quasi-single mode lasing in the high-power regime, where the device proved to be robust and stable. \\
In summary, the desired modification of nanowire-based lasers represents a promising tool for the realization of next-generation plasmonic circuits. In this work, we demonstrated that the coupling of ZnO nanowires with “fenced” Al gratings, enables the tailoring of their lasing emission. Namely, for an orientation of the nanowire perpendicular to the grating ridges (trenches), the experimental and theoretical results showed that the lasing action is dominated by a hybrid mode which is supported by the effective coupling with the localized plasmon polaritons at the fence edges. This enables a distributed-feedback-like behavior for the grating-supported resonator, which provides additional feedback to the nanowire cavity, with reflectance values as high as 50\%. With this work, we intend to foster efforts toward the optimization of tailoring schemes for such a class of nanolasers, given the ultimate goal of attaining a stable room-temperature operation, which, as of now, seems to be still highly hindered by the strong plasmonic losses. Nonetheless, the compactness of the grating structures paves the way to the integration of nanowire–metal grating arrays, for which, further mode selection schemes can be accordingly tailored and realized.

\vspace{0.5cm}

See the supplementary material for more information about the fabrication process, the nanowire manipulation procedure, additional measurements and simulation details.

\section*{AUTHOR DECLARATIONS}

\subsection*{Acknowledgments}

The authors acknowledge the financial support via the projects C5, A2, and Z3 in the CRC 1375 “NOA–Nonlinear optics down to atomic scales” funded by the Deutsche Forschungsgemeinschaft (DFG) under project-ID 398816777.

\subsection*{Conflict of Interest}

The authors have no conflicts to disclose.

\subsection*{Author Contributions}

\textbf{Francesco Vitale}: conceptualization (equal); investigation – experimental (lead); writing – original draft (lead). \textbf{Daniel Repp}: formal analysis, software – simulations (lead); writing – review \& editing (equal). \textbf{Thomas Siefke}: resources –  sample fabrication (lead); writing – review \& editing (equal). \textbf{Uwe Zeitner}: writing – review \& editing (equal). \textbf{Ulf Peschel}: writing – review \& editing (equal). \textbf{Thomas Pertsch}: funding acquisition (equal); writing – review \& editing (equal). \textbf{Carsten Ronning}: funding acquisition (equal); conceptualization (equal); project administration (lead); supervision (lead); writing – review \& editing (lead).

\bibliography{manuscript}

\end{document}


\maketitle

\section{Dielectric/metal substrate fabrication and EBL patterning}
\begin{figure} [ht!]
\centering
    \includegraphics[width=\textwidth]{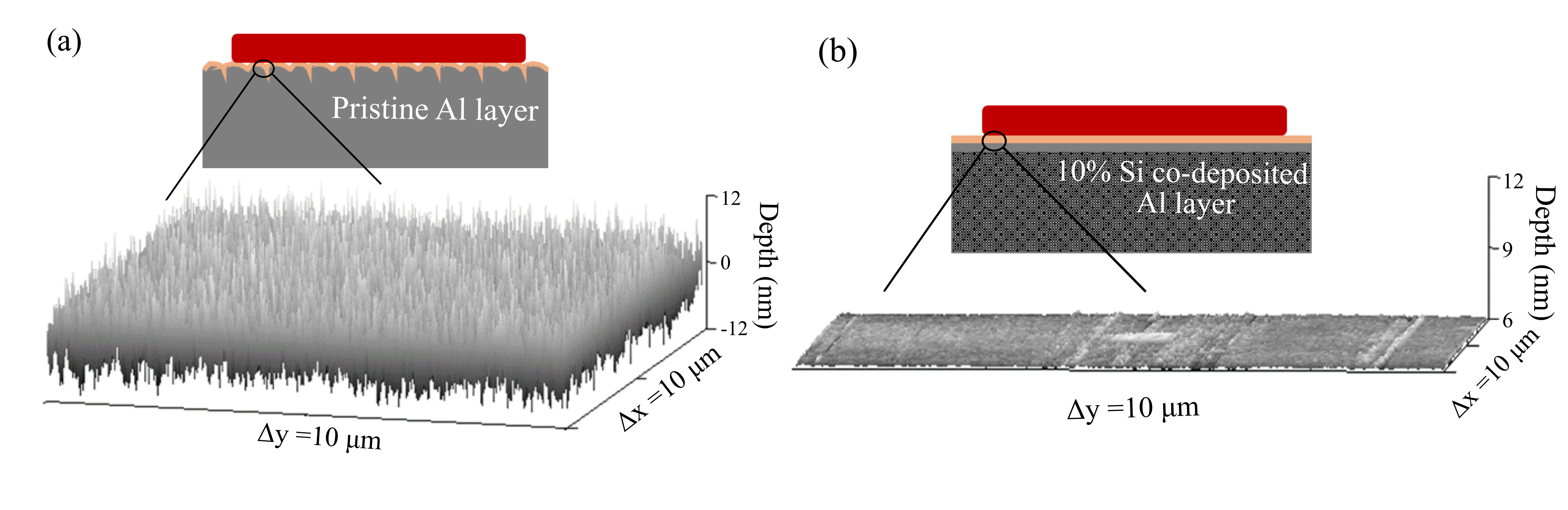}
        \caption{AFM height profiles for (a) pristine and (b) 10\% Si co-deposited polycrystalline Al substrates. The analysis revealed an average RMS surface roughness $R_q  \approx$ 3.8 nm for the pristine Al layer and  $R_q  \approx$ 0.2 nm for the co-deposited one.
}
        \label{Fig.S1}
\end{figure}

The dielectric/metal substrates were fabricated by first depositing aluminum films with a thickness of $t \approx$ 68 nm on a fused silica substrate via ion beam deposition (Ionfab 300LC+ by Oxford Instruments). High relative densities and good adhesion were achieved with this method. However, the high normalized energy flux and the generalized temperature during the film growth fostered diffusion processes and, hence, recrystallization \cite{Anders}. Therefore, materials with a low melting point, such as aluminum, tend to become polycrystalline (PC) with very large grains, leading to high surface roughness. Figure \ref{Fig.S1}(a) shows an AFM image of an as-deposited Al layer: the RMS surface roughness was found to be $R_q \approx$ 3.8 nm, calculated from measurements conducted in tapping mode over a $10 \upmu \textup{m} \times 10 \upmu \textup{m}$ area with a resolution of 40 nm. This resulted in a decreased propagation length and increased scattering for the surface plasmon polaritons at the spacer/nanowire interface \cite{Raether}. In order to reduce the roughness, 10 at\% silicon was co-deposited by switching between an aluminum and a silicon target. The deposition time was 25 s at an Ar-ion energy of 1.3 keV and a current of 250 mA for aluminum, and 10 s at an Ar-ion energy of 1.3 keV and a current of 100 mA for silicon respectively. The cycles were repeated 13 times and ended with a Si-free termination layer of thickness $t \approx$ 5-6 nm. On top of this termination layer, a native aluminum oxide layer formed with an estimated thickness of $t \approx$ 2-3 nm. 
The co-deposition with silicon led to the reduction of the surface roughness by about one order of magnitude down to $R_q \approx$ 0.2 nm: the corresponding AFM height profile is shown in Figure \ref{Fig.S1}(b).

\begin{figure} [ht!]
\centering
    \includegraphics[width=\textwidth]{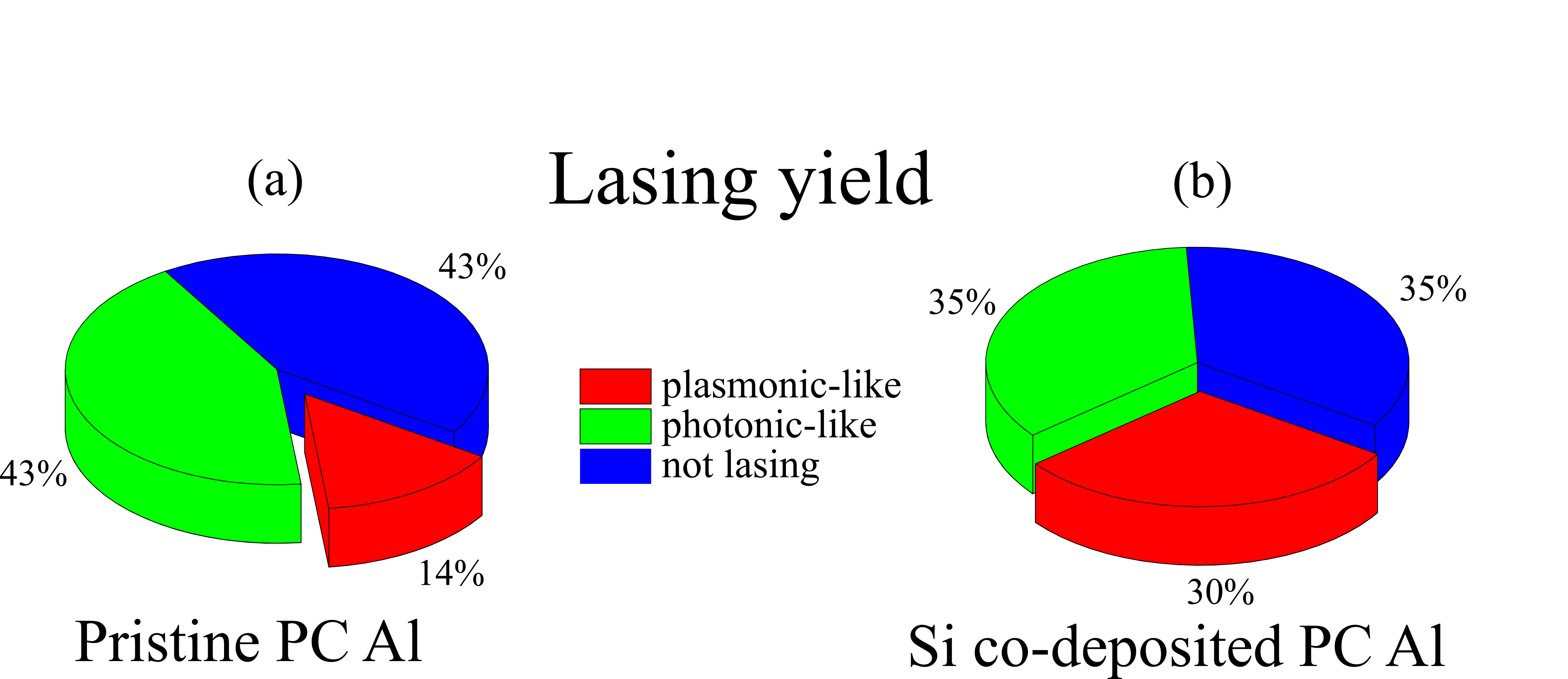}
        \caption{Lasing yield for nanowires on top of (a) pristine and (b) Si co-deposited PC aluminum layers. The improvement of the surface roughness led to an increase in the yield of devices exhibiting plasmonic-like lasing action.
}
        \label{Fig.S2}
\end{figure}

 Figure \ref{Fig.S2} shows the respective fractions of lasing yield for a sample of 30 nanowires on top of Al substrates with and without the Si co-deposition. As a consequence of the smoother surface, the yield of nanowires that showed plasmonic-like lasing at room temperature, significantly increased. This can be ascribed to the larger propagation length and reduced scattering of the surface plasmon polaritons at the spacer/nanowire interface \cite{Raether}, together with an overall better thermal dissipation that increased the optical damage threshold.

Subsequently, the grating structures were realized by character projection electron beam lithography (Vistec SB350 OS) \cite{Kley}. The mask, consisting of a positive tone CAR FEP171 (Fujifilm) was transferred onto the aluminum film by ion beam etching (Ionfab 300LC+, Oxford Instruments). The resist was, then, stripped by reactive ion etching in an oxygen atmosphere (Sentech SI-591), and, ultimately, the samples were laser diced.

\section{Sample preparation and single-nanowire nano-manipulation}

\begin{figure} [ht!]
\centering
    \includegraphics[width=\textwidth]{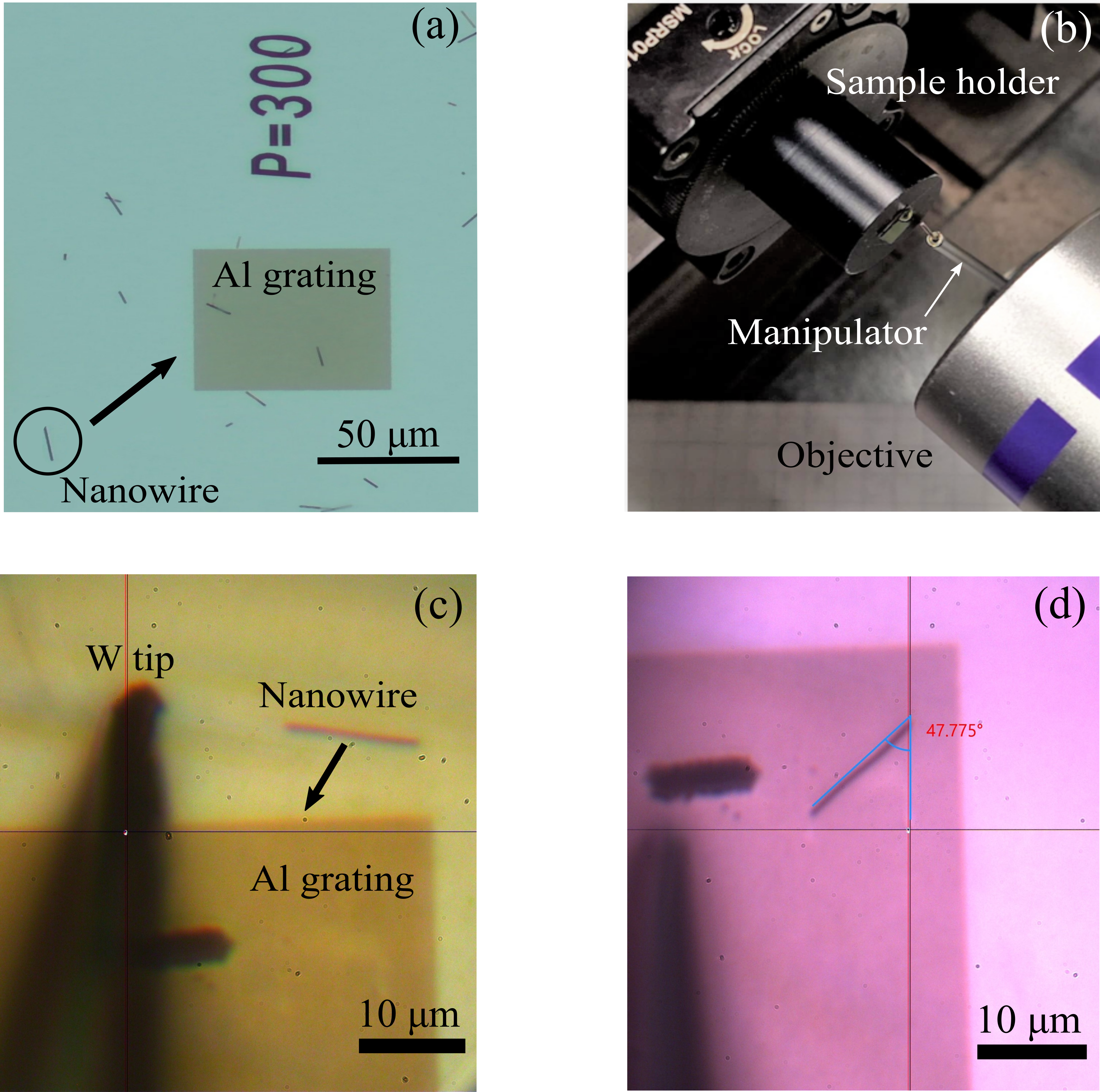}
        \caption{(a) Optical microscope image of a section of the Al$_2$O$_3$/Al substrate, exhibiting an exemplary EBL-patterned metallic grating. (b) Photograph of the nano-manipulation setup: the arrow points to the manipulator arm, equipped with the whisker mounting a W tip. (c) Representative optical image of a single nanowire and the W tip, acquired prior to the nano-manipulation. The same nanowire was successfully placed on top of the grating with a ridge (trench) orientation of $\theta \approx 50\degree$, as shown in (d). 
}
        \label{Fig.S3}
\end{figure}

The ZnO nanowires were dispersed onto the EBL-patterned Al$_2$O$_3$/Al substrates, shown in Figure \ref{Fig.S3}(a), via a dry imprint technique consisting in gently rubbing the growth sample surface against that of the target substrate, which was, in turn, imaged under an optical microscope (Zeiss Axiolab 5). The mechanical nano-manipulation was performed in air under the PL microscope, as shown in the photograph of \ref{Fig.S3}(b), in order to avoid the modification of the optical properties induced by the electron beam irradiation, as discussed in the main text of the manuscript. To this end, we made use of a remotely controlled nano-manipulator (Kleindiek MM3A-EM) featuring a W tip with a radius of curvature of about 1 micron installed in between the sample and the 50x NUV objective. The sample was imaged via a white light fiber illuminator (Thorlabs OSL1-EC) on a microscopy camera (PixeLink PL-B776U). After the tip was brought in contact with the substrate, finer displacement steps were achieved via a remotely controlled piezo-controller (Thorlabs BPC303) connected to the piezo-actuators of the sample stage. This allowed for an accurate “nudging” of the single nanowire toward the metallic grating and a successful placement on top of it, as shown in Figures \ref{Fig.S3}(c)(d), in order to avoid abrupt and uncontrolled movements of the W tip, eventually leading to the nanowire breakage.

\section{Further optical measurements of nanowires on gratings}

\begin{figure} [ht!]
\centering
    \includegraphics[width=\textwidth]{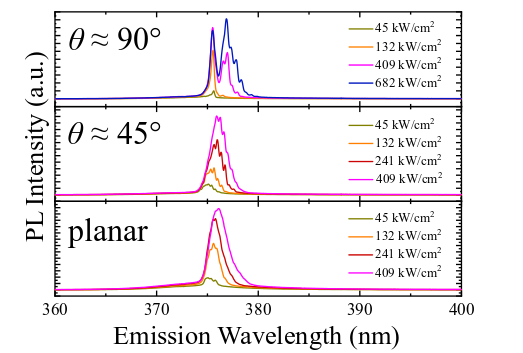}
        \caption{PL spectra of a ZnO nanowire with $L \approx$  13 $\upmu$m and $\langle d \rangle \approx$ ≈ 230 nm placed on a grating with $p$ = 300 nm and $FF$ = 0.5 at different configurations.
}
        \label{Fig.S4}
\end{figure}

Figure \ref{Fig.S4} shows the PL spectra from another nanowire featuring a length $L \approx$ 13 $\upmu$m and average diameter $\langle d \rangle \approx$ 230 nm, placed on a grating with pitch $p$ = 300 nm and fill factor $FF = 0.5$. Also in this case, we can clearly observe the emergence of additional peaks in the low-energy side of the gain envelope, for an orientation of the nanowire perpendicular to the metallic ridges. The additional peaks disappeared for a subsequent re-orientation of the wire at $\theta \approx 45\degree$ as well as for the planar situation. We attribute the slight shift ($\approx$ 1 nm) of the spectral location of the long-wavelength peak, compared to the other two nanowires discussed in the main text, to the variations related to the cavity imperfections (both nanowire- and grating-related).

\section{Active lasing simulations}

The active lasing simulations were accomplished by choosing the 4-level-2-electron material model provided by Lumerical for the nanowire material. A non-dispersive dielectric with an index of $n$ = 2.4 was chosen as a base material. The nanowire was illuminated by a Total-Field Scattered-Field (TFSF) source, modeling the experimental pump, for an excitation time $\tau_{\textup{exc}}$ = 500 fs at a wavelength $\lambda_{\textup{exc}}$ = 350 nm, in resonance with the pump transition in the material. The pump was, then, switched off, leading to the excitation of temporal Fourier components resonant with the lasing transition, thereby achieving a broad-band seeding of the lasing modes. This resulted in the appearance of a lasing pulse, which experienced gain. The lasing wavelength was identified via the observation of a maximum in the spectral power distribution from a monitor placed perpendicularly to the nanowire axis in a transversal cross-section. The field distribution was given by a second monitor plane, spanned by the nanowire axis and the surface normal to the substrate.